\documentclass[%
 reprint,
 amsmath,amssymb,
 aps,
]{revtex4-1}
\usepackage{feynmp-auto}
\usepackage{array}
\usepackage{graphicx}
\usepackage{dcolumn}
\usepackage{bm}

\begin{document}

\preprint{APS/123-QED}

\title{Perturbative Yang\textendash{}Mills theory without Faddeev\textendash{}Popov ghost fields}

\author{Helmuth Huffel}
 
\author{Danijel Markovic}%
\affiliation{%
 Faculty of Physics, University of Vienna, Boltzmanngasse 5, A-1090 Vienna, Austria.
}%

\date{\today}

\begin{abstract}
A modified Faddeev-Popov path integral density for the quantization
of Yang-Mills theory in the Feynman gauge is discussed, where contributions of the
Faddeev-Popov ghost fields are replaced by
multi-point gauge field interactions. An explicit
calculation to $O(g^{2})$ shows the equivalence of the usual Faddeev-Popov
scheme and its modified version.

\end{abstract}

\pacs{Valid PACS appear here}
\maketitle


\section{Introduction}

Faddeev and Popov \cite{Fadd} proposed a highly acclaimed path integral
quantization procedure for Yang\textendash{}Mills theory. 
Yang\textendash{}Mills theory is a gauge theory based on compact simple Lie groups. It forms the basis of our understanding of the Standard Model of particle physics \cite{Glashow,Weinberg,Salam}, which has two basic components: The spontaneously broken $SU(2) \times U(1)$ electroweak theory, and the unbroken SU(3) color gauge theory, known as Quantum Chromodynamics (QCD).\\ \\
Although electromagnetism and the weak interactions (responsible for the forces between sub-atomic particles that cause  radioactive decay) appear quite different at everyday low energies, the Standard Model understands them as two different aspects of the same force. The Higgs mechanism \cite{Higgs1,Englert,Guralnik,Higgs2,Kibble} provides an explanation for the presence of  massive gauge bosons (the  carriers of the weak force) aside of the massless photon (the carrier of the electromagnetic force).  The discoveries of the massive $W^\pm$ and $Z$ gauge bosons at the CERN \=pp collider \cite{WUA1,WUA2,ZUA1,ZUA2}  as well as of the Higgs particle at the Large Hadron Collider  \cite{ATLAS,CMS1} are considered as major successes for  the European Organization for Nuclear Research.
\\
\\
QCD is the theory which describes the strong interactions  between massive quarks and massless gluons (responsible for binding neutrons and protons to create atomic nuclei).   QCD exhibits two main properties,   asymptotic freedom \cite{Gross Wilczek,Politzer}  and color confinement. Asymptotic freedom refers to the weakness of  the strong interactions at short-distances (or high energies, respectively). It allows a perturbative treatment, which is often referred to as perturbative QCD.  Hereby high energy hadronic processes involving a large momentum transfer  can be factorized  into one part  which requires detailed nonperturbative information on parton distribution functions and into a second part, which is calculable using perturbation theory.   Parton distribution functions (specifying how hadrons are built out of quarks and gluons) have to be extracted from data and are available from various groups worldwide. The  perturbative part of the calculation is done in an expansion in the coupling constant. In this context, we  especially mention results at next-to-leading order  \cite{1412.5157}, next-to-next-to-leading order  \cite{1302.6216,1303.6254,1404.7116,1504.07922}, even next-to-next-to-next-to-leading order  \cite{1503.06056,1411.3007}, as well as calculations supplemented with resummations of  logarithmic contributions  \cite{1411.6633, 1006.3080}. 
Confinement is the  phenomenon  of non-observation of color charged particles like free quarks or gluons and is believed to follow from the strength of the QCD coupling constant at long distances (or low energies, respectively).  It should be remarked, however, that presently there is no  analytic proof of color confinement  in Yang-Mills  theory. Confinement is crucial for explaining why nuclear forces are short ranged while massless gluon exchange would be long ranged: Nucleons are colorless so they  cannot
exchange colored gluons but only colorless states. The lightest such 
particles are pions, which fixes the range of nuclear forces  by the inverse of their mass to about $10^{-14}$cm. \\
 \\
Upon quantizing Yang-Mills theory new fields
are introduced, called Faddeev-Popov ghost fields, which are associated
to the gauge fixing. Mathematically,  
 ghost fields allow for an integral representation of the Faddeev-Popov determinant (see below) in terms of a local action functional. The ghost fields 
are  Lorentz scalars but obey Fermi  statistics, they are arising inside Feynman diagrams in closed loops only. The ghosts'  unphysical degrees of freedom are needed to exactly cancel unphysical polarization states of the gauge field, leading to a unitary theory. The proof of  unitarity relies on the Slavnov-Taylor identities  \cite{Taylor,Slavnov}, which in turn  play a key role in the proof of the renormalizability  \cite{’t Hooft,’t HooftVeltman} of Yang-Mills theories. The Slavnov-Taylor identities  led Becchi, Rouet and Stora   \cite{Becchi1,Becchi2}  and, independently, Tyutin \cite{Tyutin} to discover a global supersymmetry invariance of the gauge fixed Yang-Mills action  including the ghost contributions. 
 \\
\\
The perturbation theory  of Yang\textendash{}Mills theory as developed by Faddeev-Popov,  the property of asymptotic
freedom and  renormalizability are at the heart of the Standard Model of elementary particle physics. \\
\\
In this paper a modified Faddeev-Popov path integral  quantization
of Yang-Mills theory is presented, where contributions of the
Faddeev-Popov ghost fields are replaced by
multi-point gauge field interactions.  This is a new formulation of quantum Yang-Mills theory without the use of Grassman-valued fields. 

\section{Local features of Yang-Mills theory}

Let ${\cal A}$
be the space of Yang-Mills fields and  ${\cal G}$
 the gauge group (for a detailed mathematical account of the involved space we refer to  \cite{singer,Narasimhan,Mitter Viallet}). Then ${\cal G}$ defines a principal ${\cal G}$-bundle
$\ensuremath{\ensuremath{{\cal A}\stackrel{\pi}{\longrightarrow}{\cal A}/{\cal G}=:{\cal M}}}$
over the space ${\cal M}$ of all inequivalent gauge potentials with
projection $\pi$. ${\cal M}$ represents the true degrees of freedom. However, the principal  ${\cal G}$-bundle
${\cal A\rightarrow{\cal M}}$ is not globally trivializable \cite{singer,Narasimhan,Mitter Viallet}, giving rise to the so-called Gribov ambiguity \cite{gribov}  (for a recent review see \cite{Vandersickel}).\\
 \\
It is advantageous to separate the Yang-Mills fields $A_{\mu}$ into
gauge independent and gauge dependent degrees of freedom. As this is only locally 
possible due to the non triviality of the bundle
${\cal A\rightarrow{\cal M}}$,  we consider the trivializable bundle
$\ensuremath{\pi^{-1}(U)\rightarrow U}$, where $U$ denotes a sufficiently
small neighborhood in ${\cal M}$. \\
 \\
 Under a gauge transformation  $\Omega\in{\cal G}$ the Yang-Mills field transforms according to
\begin{equation}
A_{\mu}^{\Omega}=\Omega\, A_{\mu}\,\Omega^{-1}-\frac{i}{g}(\partial_{\mu}\Omega)\,\Omega^{-1},
\end{equation}
where $g$ denotes the Yang-Mills coupling constant. In terms of the
local gauge fixing surface 
\begin{equation}
\Gamma=\{B_{\mu}\in\pi^{-1}(U)\,|\,\partial_{\mu}B_{\mu}=0\}
\end{equation}
all gauge fields in $\pi^{-1}(U)$ have the form $B_{\mu}^{\Omega}$,
where $B_{\mu}\in\Gamma$ and $\Omega\in{\cal G}$. Conversely, given
any $A_{\mu}\in\pi^{-1}(U)$, there exists a uniquely defined $\Omega(A)\in{\cal G}$
such that \mbox{$A_{\mu}^{\Omega(A)^{-1}}\in\Gamma$}. This explicitly means
that $\Omega(A)$ has to obey 
\begin{equation}
0=\partial_{\mu}(\Omega(A)^{-1}\, A_{\mu}\,\Omega(A))-\frac{i}{g}\partial_{\mu}((\partial_{\mu}\Omega(A)^{-1})\Omega(A)).\label{eq:fixing}
\end{equation}
This equation may be solved for $\Omega(A)$ as a formal power series in the gauge field $A_{\mu}$ \cite{boulware,zwanziger general,Lavelle,Capri},
which will be used in the next section.

\section{Perturbative {Yang-Mills theory without Faddev-Popov ghost fields}}

To begin with, let us recall the usual Faddeev-Popov formula \cite{Fadd} for calculating expectation
values of gauge invariant observables $f$
\begin{equation}
\langle f\rangle=\frac{\int\mathcal{D}A\,\delta(\partial_{\mu}A_{\mu})\text{\ det}\mathcal{\, F_{A}}\, e^{-S_{\text{inv}}[A]}\, f(A)}{\int\mathcal{D}A\,\delta(\partial_{\mu}A_{\mu})\text{\ det}\mathcal{\, F_{A}}\, e^{-S_{\text{inv}}[A]}},\label{eq:faddeev popov1}
\end{equation} 
displaying  integrations over unconstrained gauge fields $A_{\mu}\in\pi^{-1}(U)$ and  delta functions imposing the gauge fixing condition. Here $\text{det}\mathcal{\,F_{A}}=\text{det}\,\partial_{\mu}D_{\mu}(A)$ denotes the determinant of the  Faddeev-Popov operator and $D_{\mu}(A)$ is the covariant derivative with respect to $A_{\mu}$. The gauge invariant Yang-Mills action $S_{\text{inv}}[A]$ reads
\begin{equation}
S_{\text{inv}}[A]=\frac{1}{2}\int d^{d}x\,\text{Tr}\,\Bigl(F_{\mu\nu}F_{\mu\nu}\Bigr)
\end{equation}
and is defined in terms of the field strength 
\begin{equation}
F_{\mu\nu}=\partial_{\mu}A_{\nu}-\partial_{\nu}A_{\mu}-ig[A_{\mu},A_{\nu}].
\end{equation}
Note that in order to arrive at the Faddeev-Popov formula \cite{Fadd}
an infinite gauge group volume had to be cancelled between the numerator
and denumerator of the expression (\ref{eq:faddeev popov1}).\\
\\We prefer to represent the Faddeev-Popov formula (\ref{eq:faddeev popov1}) in the equivalent form
\cite{jaskolski,Asorey   Falceto}
\begin{equation}
\langle f\rangle=\frac{\int\mathcal{D}B\text{\ det}\mathcal{\, F_{B}}\, e^{-S_{\text{inv}}[B]}\, f(B)}{\int\mathcal{D}B\text{\ det}\mathcal{\, F_{B}}\, e^{-S_{\text{inv}}[B]}},\label{eq:faddeev popov}
\end{equation}
where the path integral is performed over constrained gauge fields
$B_{\mu}\in\Gamma$, and where $\text{det}\mathcal{\,F_{B}}=\text{det}\,\partial_{\mu}D_{\mu}(B)$
denotes the determinant of the  Faddeev-Popov operator with respect to $B_{\mu}$. 

Inspired by the stochastic quantization scheme \mbox{\cite{parisi,Damgaard,namiki}}
a generalization of the  Faddeev--Popov formula was proposed in
\cite{annals2}, where
\begin{equation}
\langle f\rangle=\frac{\int\mathcal{D}B\text{\,\ det}\mathcal{\, F_{B}}\, e^{-S_{\text{inv}}[B]}\, f(B)}{\int\mathcal{D}B\,\text{det\,}\mathcal{F}_{B}\, e^{-S_{\text{inv}}[B]}}\frac{\int\mathcal{D}\Omega\, e^{-S_{\mathcal{G}}[\Omega]}}{\int\mathcal{D}\Omega\, e^{-S_{\mathcal{G}}[\Omega]}}.\label{eq:our path integral}
\end{equation}
Here $S_{{\cal G}}\in C^{\infty}({\cal G})$ is an arbitrary functional
on ${\cal G}$, such that $e^{-S_{{\cal G}}}$ is integrable with respect
to the invariant measure $\mathcal{D}\Omega$ on ${\cal G}$. For  different modifications of the Faddeev--Popov formula see \cite{zwanziger general, parrinello}.
\\ 
\\
When evaluated on gauge invariant observables all additional \textit{finite}
contributions of the gauge degrees of freedom due to $S_{{\cal G}}$
cancel out, therefore the
generalized definition (\ref{eq:our path integral}) of expectation
values  equals (\ref{eq:faddeev popov}), which in turn is equivalent to the usual  Faddeev--Popov
formula 
 (\ref{eq:faddeev popov1}).\\ 
\\
It is our intention,  however, \textit{not} to  cancel these \mbox{finite} contributions,
but to transform  the fields $B_{\mu}\in\Gamma$ and $\Omega\in{\cal G}$
back into the original variables $A_{\mu}\in\pi^{-1}(U)$. In
this case the Jacobian
of the field transformation eliminates the Faddeev-Popov determinant,   so we
obtain \cite{annals2}
\begin{equation}
\langle f\rangle=\frac{\int\mathcal{D}A\, e^{-S_{\text{inv}}[A]-S_{\mathcal{G}}[\Omega(A)]}\, f(A)}{\int\mathcal{D}A\, e^{-S_{\text{inv}}[A]-S_{\mathcal{G}}[\Omega(A)]}}.\label{eq:faddeev popov generalized}
\end{equation}
Mind that now the path integral  is performed over unconstrained gauge
fields $A_{\mu}\in\pi^{-1}(U)$, similarly as in (\ref{eq:faddeev popov1}). Due to the absence of the Faddeev-Popov determinant, however,     Faddeev-Popov ghost fields are not present any longer.\\
 \\In this work we suggest to specify $\mathcal{S}_{\mathcal{G}}$ as
\begin{equation}
S_{\mathcal{G}}[\Omega(A)]=\frac{1}{g^{2}}\int d^{d}x\,\text{Tr}\,\Bigl((\partial_{\mu}\theta(A)_{\mu})(\partial_{\nu}\theta(A)_{\nu})^{\dagger}\Bigr),\label{eq:action gauge fix}
\end{equation}
 where 
\begin{equation}
\theta(A)_{\mu}=(\partial_{\mu}\Omega(A)^{-1})\,\Omega(A)
\end{equation}
 is defined in terms of $\Omega(A)$. \\
\\
To accommodate in $S_{\mathcal{G}}[\Omega(A)]$ the explicit expression
for $\Omega(A)$  is
an involved task and can only be achieved in a perturbative expansion
in the coupling constant $g$. Although $\mathcal{S}_{\mathcal{G}}[\Omega(A)]$
is depending on the gauge fields $A_{\mu}$ in a highly intricate
manner, the path integral ($\ref{eq:faddeev popov generalized}$)
itself is  performed over unconstrained gauge fields
$A_{\mu}\in\pi^{-1}(U)$. \\
\\
It will be seen that our choice for $\mathcal{S}_{\mathcal{G}}$ is
implying gauge field propagators in the Feynman gauge. In a sequel
paper we plan to study also the covariant $\xi$-gauges as well as
the limiting $\xi\rightarrow0$ case of the Landau gauge, when  multiplying
$\mathcal{S}_{\mathcal{G}}$ by the inverse of a gauge fixing parameter
$\xi$. \\
\\With the parametrization $\Omega(A)=e^{i\upsilon}$  one finds  
$\upsilon=g\,\upsilon_{1}+g^{2}\,\upsilon_{2}+\ldots$ with
\begin{equation}
\partial^{2}\upsilon_{1}=\partial_{\mu}A_{\mu}
\end{equation}
and
\begin{equation}
\partial^{2}\upsilon_{2}=i\partial_{\mu}\left(\frac{1}{2}[\upsilon_{1},\partial_{\mu}\upsilon_{1}]-[\upsilon_{1},A_{\mu}]\right).
\end{equation}
Correspondingly, one obtains the contributions to the gauge fixing action
$\mathcal{S}_{\mathcal{G}}=\mathcal{S}_{\mathcal{G}}^{0}+\mathcal{S}_{\mathcal{G}}^{1}+\mathcal{S}_{\mathcal{G}}^{2}+\ldots$
in a power series expansion of the coupling constant $g$. \\
 \\To lowest
order we have
\begin{equation}
\mathcal{S}_{\mathcal{G}}^{0}=\int d^{d}x\,\text{Tr}\,\Bigl((\partial_{\mu}A_{\mu})^{2}\Bigr),\label{eq:standard gauge fix}
\end{equation}
which, as advocated, represents the standard gauge fixing term of Yang-Mills theory
in the Feynman gauge. 
This standard  gauge fixing term,
however, is accompanied by additional, unconventional interaction
terms of the gauge field. To the first order in $g$ a new triplic
gauge field interaction term arises
\begin{equation}
\mathcal{S}_{\mathcal{G}}^{1}=-ig\int d^{d}x\,\text{Tr}\,\Bigl( (\partial_{\mu}A_{\mu})\,\partial_{\nu}[v_{1},A_{\nu}]\Bigr)+h.c.\label{eq:triplic},
\end{equation}
whereas the second order expansion in $g$ provides us with new quartic
gauge field interaction terms
\begin{align}
\mathcal{S}_{\mathcal{G}}^{2}= & -ig^{2}\int d^{d}x\,\text{Tr}\:\Bigl((\partial_{\mu}A_{\mu})\,\partial_{\nu}[v_{2},A_{\nu}]\Bigr)\nonumber \\
 & -\frac{1}{2}g^{2}\int d^{d}x\,\text{Tr}\,\Bigl(\left(\partial_{\mu}[v_{1},A_{\mu}]\right)\,\left(\partial_{\nu}[v_{1},A_{\nu}]\right)\Bigr)\nonumber \\
 & -\frac{1}{2}g^{2}\int d^{d}x\,\text{Tr}\,\Bigl((\partial_{\mu}A_{\mu})\,\partial_{\nu}[v_{1},[v_{1},A_{\nu}]]\Bigr)+h.c.\label{quartic}
\end{align}
$\Omega(A)$,
as well as the   gauge fixing action $\mathcal{S}_{\mathcal{G}}[\Omega(A)]$ defined in (\ref{eq:action gauge fix}),
can in principle be calculated in perturbation theory to any desired order in $g$, implying
higher and higher multi-point gauge field interaction terms. Feynman
rules corresponding to these interactions may be derived
  in addition
to the standard three-point and four-point Yang-Mills gauge field terms. 
\\
 \\
When calculating expectation values of gauge invariant observables
the new interaction terms will generate the usual Faddeev-Popov ghost contributions
order by order in perturbation theory. To explicitly verify this general claim we choose
$F_{\mu\nu}F_{\mu\nu}$ as gauge invariant observable.  Contributions  to $\mathcal{O}(g^2)$ arising from (\ref{eq:triplic}) and (\ref{quartic}) read 
\begin{equation}
\resizebox{.9\hsize}{!}{$\langle  F^2 \rangle_{new}=  (gf^{abc})^2\int \frac{d^{d}p}{(2\pi)^{d}}\frac{d^{d}q}{(2\pi)^{d}}\frac{d^{d}r}{(2\pi)^{d}}A(p,q,r)B(p,q,r) $},
\end{equation}
where $f^{abc}$ are the structure constants and 
\begin{equation}
A(p,q,r)=(2\pi)^{d}\delta^{({d})}(p+q+r)\frac{p^2q^2-(p\cdot q)^2}{p^2q^2r^2}.
\end{equation} 
The subscript "new" refers  to contributions arising only from the new three- and four-point gauge field interaction vertices. The function $B(p,q,r)$ singles out the different contributions from the various diagrams, see Table I. 
\begin {table}[h]
\label{Tab.1}
\begin{center}
 \begin{tabular}{|>{\centering\arraybackslash}m{3cm}|>{\centering\arraybackslash}m{3cm}|} 
      \hline
      \vspace{5pt}
     \begin{fmffile}{diagram1}
      \fmfset{thin}{.5pt}
      \fmfset{wiggly_len}{2mm}
      \begin{fmfgraph*}(50,30)
	\fmfleft{i}
	\fmfright{o}
	\fmf{wiggly,tension=2}{i,v1}
	\fmf{wiggly,tension=2}{v2,o}
	\fmf{wiggly,right}{v1,v2,v1}
	\fmfv{decor.shape=circle,filled=full,decor.size=5}{v1}
      \end{fmfgraph*}
      \end{fmffile}
      &
      $ -\frac{2}{r^2} $\tabularnewline \hline
      \vspace{5pt}
       \begin{fmffile}{diagram2}
      \fmfset{thin}{.5pt}
      \fmfset{wiggly_len}{2mm}
      \begin{fmfgraph*}(50,30)
	\fmfleft{i}
	\fmfright{o}
	\fmf{wiggly,tension=2}{i,v1}
	\fmf{wiggly,tension=2}{v2,o}
	\fmf{wiggly,right}{v1,v2,v1}
	\fmfv{decor.shape=circle,filled=full,decor.size=5}{v1}
	\fmfv{decor.shape=circle,filled=full,decor.size=5}{v2}
      \end{fmfgraph*}
      \end{fmffile}
      &
      $ \frac{q^2+r^2}{p^2r^2} $\tabularnewline \hline
      \vspace{5pt}
       \begin{fmffile}{diagram3}
      \fmfset{thin}{.5pt}
      \fmfset{wiggly_len}{2mm}
      \begin{fmfgraph*}(50,25)
    \fmfbottom{i,o}
    \fmftop{t}
    \fmf{wiggly}{i,v1,o}
    \fmffreeze
    \fmf{wiggly,right,tension=1}{v1,t,v1}
    \fmfv{decor.shape=square,filled=full,decor.size=5}{v1}
      \end{fmfgraph*}
      \end{fmffile}
      &
      $ \frac{2p^2-q^2}{p^2r^2} $\tabularnewline \hline
      \vspace{5pt}
     \begin{fmffile}{diagram4}
      \fmfset{thin}{.5pt}
      \fmfset{wiggly_len}{2mm}
      \begin{fmfgraph*}(30,30)
	\fmfleft{i1,i2}
	\fmfright{o}
	\fmf{wiggly}{i1,v1,o}
	\fmf{wiggly}{i2,v1}
	\fmfv{decor.shape=circle,filled=full,decor.size=5}{v1}
      \end{fmfgraph*}
      \end{fmffile}
      &
      $ -\frac{2}{p^2} $\tabularnewline \hline\hline
      \vspace{5pt}
       \begin{fmffile}{diagram5}
      \fmfset{dot_len}{.6mm}
      \fmfset{thin}{.5pt}
      \fmfset{wiggly_len}{2mm}
      \begin{fmfgraph*}(50,30)
	\fmfleft{i}
	\fmfright{o}
	\fmf{wiggly,tension=2}{i,v1}
	\fmf{wiggly,tension=2}{v2,o}
	\fmf{dots,right}{v1,v2,v1}
      \end{fmfgraph*}
      \end{fmffile}
      &
      $ -\frac{1}{p^2} $ \tabularnewline \hline 
\end{tabular}
\caption {Contributions to $B(p,q,r)$ \\ \\The solid dot represents the new three-point gauge field interaction vertex and the solid square the new four-point gauge field interaction vertex. Vertices without any special labeling correspond to the standard Yang-Mills ones.  The wiggly gauge field lines represent propagators in the Feynman gauge. The bottom line of the table displays the standard Faddeev-Popov ghost field contribution.}
\end{center}
\end{table}
\\
 \\
Summing up all  terms  precisely gives the  usual  contribution of the  Faddeev-Popov ghost fields. We  employed Mathematica as supplement to the involved calculations
done by hand. Details of the procedure, including the new Feynman
rules, will be reported elsewhere.
\section{Outlook}

Perturbative Yang\textendash{}Mills theory without \mbox{Faddeev}\textendash{}Popov ghost fields has been presented in this paper and shown to be viable and useful. We are confident that various generalizations and a new arena of promising applications will open up.
\\ \\
First we plan  doing perturbative calculations for various sets of observables. The efficiency of our method   and its general  performance in comparison   to the  conventional Faddev-Popov procedure will be studied. 
\\ \\
We expect to extend our method     for calculations in the covariant $\xi$-gauge as well as for the limiting   $\xi\rightarrow 0$ case of the Landau gauge.
\\ \\
In  another approach we envisage to  adapt  our scheme for covariant gauge-fixing procedures in lattice gauge theories \cite{ Fachin1, Fachin2,parrinello2,Maas1,Maas2,Cucchieri2}, for a review see \cite{Maas}. 
\\ \\
A  further challenge will be to formulate  generalized   Slavnov-Taylor identities  \cite{Taylor,Slavnov} and set up a perturbative renormalization program.
\\ \\
Finally we propose to discuss  our method beyond its local perturbative validity, addressing  the  issue of Gribov ambiguities. Similar to
 \cite{gribov,zwanziger local,zwanziger renorm} the functional
integration  could be restricted to a subset of the gauge field space. Alternatively we  suggest  the possibility of partitioning the whole space of
gauge fields into patches - where gauge fixing without Gribov ambiguity is possible locally
 - and summing appropriately over all patches
\cite{global}.

\section*{Acknowledgments }

We thank Gerald Kelnhofer and Axel Maas for valuable
discussions and Maximilian L\"oschner for providing us with the figures.  The present work initiated from \cite{annals2} and H.H. is especially indebted to Gerald Kelnhofer for encouragement and support.

\end{document}